\documentclass[12pt]{article}

\usepackage{amsmath}
\usepackage{amssymb}

\newtheorem{theorem}{Theorem}[section]
\newtheorem{corollary}{Corollary}[section]
\newtheorem{prop}{Proposition}[section]
\newtheorem{definition}{Definition}[section]
\numberwithin{equation}{section}

\begin{document}

\begin{center}
{\bf Toeplitz Quantization without} \\ {\bf Measure or Inner Product}
\end{center}

\centerline{Stephen Bruce Sontz}

\begin{center}
Centro de Investigaci\'on en Matem\'aticas, A. C.
\\
 (CIMAT)
   \\
Guanajuato, Mexico
\vskip 0.4cm
email: sontz@cimat.mx
\end{center}

\begin{abstract}
   This note is a follow-up to a recent paper by the author.
   Most of that theory is now realized in a new setting where the vector space of
   symbols is not necessarily an algebra nor is it equipped with an inner product,
   although it does have a conjugation.
   As in the previous paper one does not need to put a measure on this vector space.
   A Toeplitz quantization is defined and shown to have most
   of the properties as in the previous paper, including creation and
   annihilation operators. 
   As in the previous paper
   this theory is implemented
   by densely defined Toeplitz operators which act in a Hilbert space,
   where there is an inner product,  of course.
   Planck's constant also plays a role in the canonical commutation relations of 
   this theory.
\end{abstract}

\linebreak
creation operators.

\section{Introduction}
In a recent paper \cite{tqac} of mine I developed a theory
of a Toeplitz quantization whose symbols lie in a possibly
non-commutative algebra which has an inner product.
At that time I was motivated by previous papers (\cite{part2} and \cite{quantum-plane})  of mine that
had symbols in a non-commutative algebra. In those cases
there was also an inner product available which served more
than anything as a part of a formula defining a projection operator.
And that projection operator was used in the standard way to define
Toeplitz operators in that setting. But now I have realized that there is
another way to arrive at most of the results of \cite{tqac} without supposing
that the complex vector space (no longer assumed to be an algebra) of symbols has an inner product,
though I still require that it have a conjugation to get more interesting results.

While the paper \cite{tqac} presented a viable quantization scheme that did not
involve a measure, the objection could be made that an inner product is some sort
of mild generalization of a measure, that it is a `measure in disguise' or some such criticism.
However, in this note there is neither measure nor inner product on the `classical' space of symbols.  
Of necessity there is an inner product in the quantum Hilbert space.

The references for this short note are deliberately kept to just a very few. 
For further background and motivation
on this topic see \cite{tqac}, consult the references found there
and continue recursively.

\section{The new setting}
We have a new setting that has some things in common with that in \cite{tqac}.
So, to facilitate this presentation I will use 
the same notation as in \cite{tqac}.
Here are the exact structures to be considered in this note together with their notations.
They involve three vectors spaces (denoted by $\mathcal{A}$,  $\mathcal{H}$ and
 $\mathcal{P}$) over the field $\mathbb{C}$ of complex numbers.
 These spaces are required to satisfy these eight conditions:
 
\begin{enumerate}

\item $\mathcal{H}$ is a Hilbert space. 

\item $\mathcal{A}$ 
has a conjugation denoted by $g^*$ for all $g \in \mathcal{A}$.
A conjugation is by definition  an anti-linear, involutive mapping of a vector space to itself.

\item $\mathcal{P}$ is a dense subspace of $\mathcal{H}$.

\item $\mathcal{P}$ is a vector subspace of $\mathcal{A}$.

\item  $\mathcal{P}$ is an associative algebra with unit $1$ satisfying $1^* = 1$.
Note that $\mathcal{P}$ is not necessarily commutative.

\item There is a left action of $ \mathcal{P}$ on $ \mathcal{A} $.
This means that there
is a unital algebra morphism $\mathcal{P} \to \mathrm{End} ( \mathcal{A} )$, since
$\mathrm{End} ( \mathcal{A} )$ acts by convention on the left of $\mathcal{A}$.
In particular we assume that this action 
(thought of as a bilinear map $ \mathcal{P} \times \mathcal{A} \to \mathcal{A}$) restricts to the multiplication
map $ \mathcal{P} \times \mathcal{P} \to \mathcal{P}$ of the algebra $\mathcal{P}$.
The notation is $ (\phi, g) \mapsto \phi g$ for $ (\phi, g) \in \mathcal{P} \times \mathcal{A} $.

\item There is a linear map $P : \mathcal{A} \to \mathcal{P} \subset \mathcal{A}$ which
satisfies $P^2 = P$ and with range $\mathrm{Ran} \, P = \mathcal{P}$.
(The co-domain of $P$ is taken to be either $\mathcal{P}$ or $\mathcal{A}$, as convenience dictates.)
One immediately has that the restriction of $P$ to $\mathcal{P}$ is the identity map on $\mathcal{P}$.

\item
$ \langle T_g \phi_1, \phi_2 \rangle_{\mathcal{H} } = \langle \phi_1, T_{g^*} \phi_2 \rangle_{\mathcal{H} }
   \, \mathrm{for~all~} \phi_1, \phi_2 \in \mathcal{P}$ 
  and $g \in \mathcal{A}$ where  $T_g$, the Toeplitz operator with symbol $g$, will be defined below.
  This condition means $T_{g^*} \subset (T_g)^*$, the adjoint of $T_g$.

\end{enumerate}

I do not assume that there is an inner product on $\mathcal{A}$, but of course we do
have an inner product, denoted by $\langle \cdot, \cdot \rangle_{\mathcal{H} }$, 
on the Hilbert space $\mathcal{H}$.
And this restricts to an inner product on $\mathcal{P}$  thereby making it a pre-Hilbert space.
In \cite{tqac} the vector space $\mathcal{A}$ of symbols was assumed to be an algebra.
We retain the notation, but not that hypothesis, for this space. 
The conjugation on $\mathcal{A}$ typically will not leave $\mathcal{P}$ invariant.
All that we can say in general is that $ \mathcal{P}^* \subset \mathcal{A}$.
A natural way to define an inner product on $ \mathcal{P}^*$ is
$$
      \langle \psi_1, \psi_2 \rangle_{\mathcal{P}^* } :=  \langle \psi_2^*, \psi_1^* \rangle_{\mathcal{H} }
$$
for all $\psi_1, \psi_2 \in \mathcal{P}^*$.
With this inner product  $ \mathcal{P}^*$ becomes a pre-Hilbert space, which is 
anti-unitarily equivalent to $\mathcal{P}$ via the map $\phi \to \phi^*$ for all $\phi \in  \mathcal{P}^*$.
The completion of $ \mathcal{P}^*$ is denoted by $\mathcal{H}^*$.
Bearing in mind typical examples from classical analysis, one sees that $\mathcal{H}$
corresponds to a Hilbert space of holomorphic functions while $\mathcal{P}$ corresponds
to its subspace of holomorphic polynomials.
Similarly, $\mathcal{H}^*$ and $\mathcal{P}^*$ are their anti-holomorphic counterparts.
Given this intuition behind these structures, one sees that the requirement 
$\mathcal{P} \cap \mathcal{P}^*=\mathbb{C}1$ is quite natural.
However, it is not needed for the present theory, nor was it used in \cite{tqac}.
So, we will not make any assumption on $\mathcal{P} \cap \mathcal{P}^*$.

The main differences from the setting in \cite{tqac} are that $\mathcal{A}$ no longer need be an algebra
nor need it have an inner product defined on it.
However, its subspace $\mathcal{P}$ has the restriction of the inner product of $\mathcal{H}$.
Condition~6 is new in its details, but preserves the idea of the assumption as given in \cite{tqac} that $\mathcal{P}$
is a subalgebra of the algebra $\mathcal{A}$.
Condition~7 was a consequence of other assumptions given in \cite{tqac} about the existence of
a certain subset $\Phi$ of $\mathcal{P}$. 
Here it is simply taken as an additional assumption that replaces the assumptions about 
that subset $\Phi$.

In Condition~8 we require the consistency of the conjugation in $\mathcal{A}$
and the adjoint operation of operators.
In \cite{tqac} this was a consequence of an identity that itself was assumed as a hypothesis.
(See Theorem~3.3, part~4.)
Here we take this property itself itself as a hypothesis.
Of course, Toeplitz operators will be defined presently without using Condition~8.

The theory in \cite{tqac} satisfies these eight conditions.
So, the theory in this new setting generalizes the theory in \cite{tqac}.
But we see no way to define an inner product on $\mathcal{A}$ nor to extract the set $\Phi$ in this new setting.
Also, $\mathcal{A}$ in this note need not be an algebra.
So it seems safe to say that this note has a strict generalization of the theory presented in \cite{tqac}.
Nonetheless, most of the results in \cite{tqac} remain true in this new setting.

\section{Definitions and Basic Results}

We now present and prove all those results in \cite{tqac} which are still valid in this new setting.
First, here are some definitions almost identical to those in \cite{tqac}.
These are simply the natural definitions of Toeplitz operator and
Toeplitz quantization in this new setting.

\begin{definition}
For any $g \in \mathcal{A}$ define $M_g : \mathcal{P} \to \mathcal{A}$ by $M_g \phi := \phi g$
for all $\phi \in \mathcal{P}$. 
(Recall $\phi g$ is the left action of $\phi \in \mathcal{P}$ on $g \in \mathcal{A}$.)
Then define the {\em Toeplitz operator $T_g : \mathcal{P} \to \mathcal{P}$ associated to the symbol 
$g \in \mathcal{A}$} by $ T_g := P M_g$.

We let $\mathrm{End}(\mathcal{P})$ denote the vector space of all linear maps 
$\mathcal{P} \to \mathcal{P}$.
The linear map $T : \mathcal{A} \to \mathrm{End}(\mathcal{P})$ defined by 
$T : g \mapsto T_g$ is called the {\em Toeplitz quantization}.

We also consider $T_g$ as a densely defined linear operator defined in (but not on) the Hilbert
space $\mathcal{H}$ as follows:
$$
             \mathcal{P} \stackrel{T_g}{\longrightarrow} \mathcal{P} \subset \mathcal{H}.
$$
Viewed this way the domain of $T_g$ is given by $\mathrm{Dom} (T_g) = \mathcal{P}$.
\end{definition}
So each Toeplitz operator in this setting is defined in the same dense subspace 
$\mathcal{P}$,
which is invariant under the  action of $T_g$.
Consequently the composition of the Toeplitz operators $T_g$ and $T_h$ is an operator
in $\mathrm{End}(\mathcal{P})$
though it need not be itself a Toeplitz operator.
Whether a Toeplitz operator is bounded depends on more specific information
about the symbol.
Some light is already cast on these considerations by the next theorem, which is
a standard, expected result for Toeplitz operators.

\begin{theorem}
\label{theorem-3-1}
The Toeplitz quantization has the following properties:
\begin{enumerate}

\item 
$T_1 = I_{ \mathcal{P} }$, the identity map of $\mathcal{P}$.

\item $g \in \mathcal{P}$ implies that $T_g = M_g$.

\item If $g \in \mathcal{A}$ and $\psi \in \mathcal{P}$, then $T_g T_\psi = T_{\psi g}$.

\end{enumerate}

\end{theorem}
\textbf{Proof:}
We let $\phi \in \mathcal{P}$ be arbitrary throughout the proof.

For Part~1  we calculate
$
          T_1 \phi = P M_1 \phi = P (\phi 1) = P(\phi) = \phi,
$
since $P$ acts as the identity on $\mathcal{P}$.

For Part~2 we have $T_g \phi = P M_g \phi = P ( \phi g ) = \phi g = M_g \phi$,
where we used that $\phi g \in \mathcal{P}$, which follows from $\phi , \, g \in \mathcal{P}$.

For Part~3 we let $g \in \mathcal{A}$ and  $\psi \in \mathcal{P}$.
Then we calculate
\begin{align*}
        T_g T_\psi \phi &= P M_g P M_\psi \phi = P M_g (P (\phi \psi) ) = P M_g ( \phi \psi ) 
        \\
        &= P ( \phi \psi g ) = P M_{\psi g} \phi = T_{\psi g} \phi.
\end{align*}
Here we used $P (\phi \psi) =  \phi \psi$, since $\mathcal{P}$ is an algebra and so $\phi \psi \in \mathcal{P}$.
$\quad \blacksquare$

Part~1 shows that a Toeplitz operator can be bounded yet not compact.
And Part~3 shows that
the composition of two Toeplitz operators can itself be a Toeplitz operator, in which case
the symbol of the composition is given by a simple formula involving the
symbols of the factors, that is, the symbol calculus is rather straightforward in this case.

As promised Condition~8 was not used in the definition of a Toeplitz operator.
Also  Condition~8 implies that $T_g$ is a symmetric operator if 
$g$ is a self-adjoint element of $\mathcal{A}$, namely $g = g^*$.
Whether this symmetric operator has any self-adjoint extensions and, in particular,
whether it is essentially self-adjoint, are in general delicate questions that can be addressed with
functional analysis.
However, $T_1 = I_{ \mathcal{P} }$ trivially has a self-adjoint extension, namely $I_{ \mathcal{H} }$.

\begin{theorem}
Each Toeplitz operator $T_g$ is closable
and its closure, denoted by $\overline{T_g}$, satisfies
$$
     \overline{T_g} = (T_g)^{**} \subset ( T_{g^*} )^*
$$
for every $g \in \mathcal{P}$.
\end{theorem}
\textbf{Proof:}
By functional analysis 
an operator $R$ is closable if and only if $\mathrm{Dom} \, R^*$ is dense.
However $\mathrm{Dom} (T_g)^* \supset \mathrm{Dom} \, T_{g^*} = \mathcal{P}$ and $\mathcal{P}$ is
dense in $\mathcal{H}$. 
So, $\mathrm{Dom} (T_g)^*$ is itself a dense subspace and therefore $T_g$ is closable. 
Then by functional analysis $ \overline{T_g} = (T_g)^{**}$.
Finally,  $ (T_g)^{**} \subset ( T_{g^*} )^*$
follows by taking the adjoint of $T_{g^*} \subset (T_g)^*$.
(See \cite{reed-simon1} for the functional analysis results.)
$\quad \blacksquare$

Because this is a rather specific setting, one could expect a more explicit description
of the closure of a Toeplitz operator.
However, we leave this as a consideration for future research.

Theorem 3.2 in \cite{tqac} that identifies the kernel of $T$ does not go over to this setting; neither do its
consequences.
However, we can see that $g \in \ker  T$ if $g \in \mathcal{P}$ and $M_g = 0$, the zero operator.
Also, Condition~8 implies that the subspace $\ker \, T$ is closed under conjugation.
We do have the following direct consequence of the definitions, although a more computable result 
clearly would be desirable.
\begin{prop}
$g \in \ker  T$ if and only if $\mathrm{Ran} \, M_g \subset \ker \, P$.
\end{prop}

\section{Creation and Annihilation Operators}

We have creation and annihilation operators in this setting.
\begin{definition}
Let $g \in \mathcal{P}$ be given.
Then the {\em creation operator} associated to $g$ is defined to be
$$
            A^* (g) := T_g
$$
and the  {\em annihilation operator} associated to $g$ is defined to be
$$
            A (g) := T_{g^*}.
$$
\end{definition}
These are reasonable definitions since they agree with
the usual formulas for  these operators as found, for example, in \cite{quantum-plane}.
Notice that $g \mapsto A^* (g) $ is linear while $g \mapsto A (g)$ is anti-linear.
Also $ A^* (g) = T_g = M_g$ holds, because $g \in \mathcal{P}$.
Since $A^* (1) = A(1) = T_1 = I_{ \mathcal{P} }$, we see that $I_{ \mathcal{P} }$
is both a creation and an annihilation operator. 
More generally, for any $g \in \mathcal{P} \cap \mathcal{P}^*$, one has $T_g = A^* (g) = A (g^*)$ and so
$T_g$ is both a creation and an annihilation operator. 

One of the important contributions of Bargmann's seminal paper \cite{bargmann}
is that it realizes the creation and annihilation operators introduced by Fock as adjoints
of each other with respect to the inner product on the Hilbert space
which is nowadays called the Segal-Bargmann space. 
In the present setting the creation operator $A^* (g)$ and the annihilation operator $A (g)$
also have this relation, modulo domain considerations, as we have 
already seen in Condition~8.
Whether each is \textit{exactly} the adjoint of the other 
is an open question if $\mathcal{P}$ has infinite dimension, but is true
for finite dimensional $\mathcal{P}$.

In this setting, unlike that in \cite{tqac}, there is only one definition possible
for an anti-Wick quantization.
\begin{definition}
We say that $T$ is an  {\em  anti-Wick quantization} if 
$$
T_{h g^*} = T_{g^*} T_{h} 
$$
for all $g, h \in \mathcal{P}$.
Notice that $h g^*$ makes sense since it is the left action of $h \in \mathcal{P}$
on an element of $\mathcal{A}$.
\end{definition}
Notice that on the right side of this definition we have the product of an
annihilation operator $T_{g^*} $ to the left of a creation operator $T_h$.
And so the right side is in what is known as  \textit{anti-Wick order}.

In \cite{tqac} we defined $T$  an {\em alternative anti-Wick quantization} if 
the equation
$
T_{g^* h} = T_{g^*} T_{h}  
$
is satisfied
for all $g, h \in \mathcal{P}$.
But in this setting the expression $g^* h$ has not even been defined.
So this concept does not apply here.

\begin{theorem}
The Toeplitz quantization $T$ is an anti-Wick quantization.
\end{theorem}
\textbf{Proof:}
Take $g, h \in \mathcal{P}$.
Then 
$
    T_{h g^*} =  T_{g^*}  T_{h}
$,
where we have used Part~3 in Theorem \ref{theorem-3-1}.
$\quad \blacksquare$

This proof replaces the rather lengthy proofs by explicit calculations given
in \cite{part2} and \cite{quantum-plane}.

\begin{corollary}
If $\mathcal{A} = \mathcal{P} \mathcal{P}^*$, then 
one can write any Toeplitz operator as a finite sum of terms in anti-Wick order.
\end{corollary}
\textbf{Proof:}
Let $f \in \mathcal{A}$ be a symbol. 
The hypothesis means that we can write $f$ as a finite sum, 
$f = \sum_k h_k g_k^*$ with $g_k, h_k \in \mathcal{P}$, 
where $h_k g_k^*$ is the left action of $h_k \in \mathcal{P}$ on an
element of $\mathcal{A}$.
So, $T_f = \sum_k T_{g_k^*} T_{h_k}$. 
$\quad \blacksquare$

To show more clearly that our definition of anti-Wick ordering compares well with the
discussion of this topic in Theorem~8.2 in \cite{brian} we prove the next result.
But first we need a definition that is a modification for this setting of a definition
given in \cite{tqac}.

\begin{definition}
We say that $\mathcal{P}$ is {\rm $*$-friendly} 
if $\mathcal{P}^*$ is an algebra and if its multiplication satisfies
$(p_1 \cdots p_n)^* = p_n^* \cdots p_1^*$ for all $p_1 , \dots , p_n \in \mathcal{P}$.
\end{definition}
One point of this definition is that we do not require $(p_1 \cdots p_n)^*$ to be 
an element in $\mathcal{P}$.
If $\mathcal{A}$ is a $*$-algebra, then $\mathcal{P}$ is $*$-friendly where the
multiplication on $\mathcal{P}^*$ is the restriction of that on $\mathcal{A}$.

The Toeplitz quantization is a linear map whose co-domain is an algebra
and whose domain contains an algebra, namely $\mathcal{P}$.
And in the $*$-friendly case its domain also contains the algebra $\mathcal{P}^*$.

\begin{theorem}
Suppose that $g_1, \dots, g_n, h_1, \dots h_m \in \mathcal{P}$.
Then 
\begin{enumerate}

\item $T_{g_1 \cdots g_n} = T_{g_n} \cdots T_{g_1}$.

\item $T_{h_1^* \cdots h_m^*} = T_{h_m^*} \cdots T_{h_1^*}$ if
$\mathcal{P}$ is a $*$-friendly.

\item $T_{(g_1 \cdots g_n) (h_1^* \cdots h_m^*)} 
       = T_{h_m^*} \cdots T_{h_1^*} T_{g_n} \cdots T_{g_1}$ if $\mathcal{P}$ is $*$-friendly.

\end{enumerate}
\end{theorem}
\textbf{Proof:}
For Part~1 we use induction. 
The case $n=1$ is trivial, while the case $n=2$ follows from Part~3 in Theorem \ref{theorem-3-1}.
For $n \ge 3$ we have that
\begin{equation*}
T_{g_1 g_2  \cdots g_n} = T_{g_1 (g_2  \cdots g_n) } 
= T_{g_2  \cdots g_n} T_{g_1}  
= T_{g_n} \cdots T_{g_2} T_{g_1},  
\end{equation*}
where we used Part~3 in Theorem \ref{theorem-3-1}
for the second equality and 
the induction hypothesis for $n-1$ for the third equality.

For the proof of Part~2 we take the notation $T_f^*$ for any $f \in \mathcal{A}$
to mean the restriction of the adjoint  $(T_f)^*$ of $T_f$ to the algebra $\mathcal{P}$.
So, $T_f^* = T_{f^*}$ follows form Condition~8.
We then note that
\begin{equation*}
       T_{h_m^*} \!\! \cdots T_{h_1^*}  = T_{h_m}^* \!\!\! \cdots T_{h_1}^*
       = (T_{h_1} \!\! \cdots T_{h_m})^*
       = (T_{h_m \cdots h_1})^* 
       = T_{(h_m \cdots h_1)^*}
       = T_{h_1^* \cdots h_m^*}
\end{equation*}       
where we used Part~1 in the third equality and that $\mathcal{P}$ is a $*$-friendly in the
last equality.

For Part~3 we first remark that $(g_1 \cdots g_n) (h_1^* \cdots h_m^*)$ exists since it
is the left action of the element $g_1 \cdots g_n \in \mathcal{P}$ on the element
$h_1^* \cdots h_m^* \in \mathcal{P}^* \subset \mathcal{A}$. 
Then we have that
\begin{equation*}
     T_{h_m^*} \cdots T_{h_1^*} T_{g_n} \cdots T_{g_1} = 
     T_{h_1^* \cdots h_m^*} T_{g_1 \cdots g_n} 
     = T_{(g_1 \cdots g_n) (h_1^* \cdots h_m^*)}
\end{equation*}
by applying Parts~1 and 2 in the first equality and 
Part~3 of Theorem \ref{theorem-3-1} in the second equality, using $g_1 \cdots g_n \in \mathcal{P}$.
$\quad \blacksquare$

\section{Canonical Commutation Relations}

We now consider the canonical commutation relations which are satisfied by the
creation and annihilation operators.
However, our approach here is the opposite of the usual approach in which 
one starts with some deformation of the standard canonical commutation relations,
and then one looks for representations of those relations by operators
in some Hilbert space. 
Here we ask what are the appropriate canonical commutation relations that are associated
with a given Toeplitz quantization.
So, the operators acting in a Hilbert space are given first.
This section only contains definitions and a discussion of them.
It is basically the framework of a program for future research.

\begin{definition}
The subalgebra of $\mathrm{End}(\mathcal{P})$ generated by all the creation and annihilation operators
 is defined to be the {\rm algebra of canonical commutation relations}
and is denoted by $\mathcal{CCR}(P)$.
\end{definition}

We define $\mathcal{F} = \mathbb{C} \{ \mathcal{P} \cup \mathcal{P}^* \}$ to be 
the free algebra over $\mathbb{C}$ generated by
the \textit{set} $\mathcal{P} \cup \mathcal{P}^*$.
Notice that $\mathbb{C}1 \subset \mathcal{P} \cap \mathcal{P}^*$.
To avoid confusion, we will write 
the algebra generators of $\mathcal{F}$ as $G_f$ for $f \in \mathcal{P} \cup \mathcal{P}^*$.
So $\mathcal{F}$ is the complex vector space with a basis given by the monomials
$G_{f_1} G_{f_2} \cdots G_{f_n}$  of degree $n$, where $f_j \in \mathcal{P} \cup \mathcal{P}^*$
for each $j$.
We define the algebra morphism $\pi : \mathcal{F} \to \mathcal{CCR}(P)$
by $\pi ( G_f) :=  T_f$ for all $f \in \mathcal{P} \cup \mathcal{P}^*$.
Since the algebra $\mathcal{F}$ is free on the $G_f$'s, this defines $\pi$ uniquely.
Also since the elements $T_f$ for $f \in \mathcal{P} \cup \mathcal{P}^*$ are algebra
generators for the algebra $\mathcal{CCR}(P)$, 
we see that $\pi$ is an epimorphism.
We define the \textit{ideal of canonical commutation relations} in $\mathcal{F}$ 
to be $\mathcal{R}:= \ker \pi$.
Any minimal set of algebra generators of $\mathcal{R}$ is called
a \textit{set of canonical commutation relations}.
Notice that such a set will not be unique in general.

The usual canonical quantum mechanical
 commutation relations (when written as ideal generators
given by $a_j a_k^* - a_k^* a_j - \hbar \delta_{j,k} 1$) have the property that 
for $j \ne k$ they are homogeneous in the variables $a_j$ and $a_k^*$
and do not include any quantum effect due to Planck's constant $\hbar$. 
In this case they correspond to the commutativity of classical mechanical variables.
However, for $j=k$ they are not homogeneous in the variables, and they do include $\hbar$.
Moreover, in this case the classical relation is obtained by dropping the
lower order `quantum correction'.
These remarks motivate the following definition.

\begin{definition}
We say that a homogeneous element in $\mathcal{R} \subset \mathcal{F}$ is a {\rm classical relation}
and that a non-homogeneous element in $\mathcal{R}$ is a {\rm quantum relation}.

Suppose $R \in \mathcal{R}$ is a non-zero relation. 
Then we can write $R$ uniquely as $R = R_0 + R_1 + \cdots + R_n$,
where each $R_j$ is homogeneous with
 $\deg \, R_j = j$ for each $j = 0,1, \dots, n$ and $R_n \ne 0$.
Then we say that $R_n$ is the {\rm classical relation associated to~$R$}.
\end{definition}
Of course, $R_n$ is actually a classical relation.
Both of the cases $R_n \in \mathcal{R}$ and $R_n \notin \mathcal{R}$ can occur
as the example before this definition shows.
What we are doing
intuitively to get the classical relation $R_n$ from $R$ is to discard the `quantum corrections' 
$R_0, R_1, \dots, R_{n-1}$ in $R$.
We next define
$$
\mathcal{R}_{cl}:= \langle R_n \, | \, R_n 
\mathrm{~is~the~classical~relation~associated~to~some~}
R \in \mathcal{R} \rangle,
$$
where the brackets $\langle \cdot \rangle$ indicate that we are taking the
two-sided ideal in $\mathcal{F}$ generated by the elements inside the brackets.
\begin{definition}
The {\rm dequantized algebra} associated to $\mathcal{A}$ is defined to be
$$
\mathcal{DQ}:= \mathcal{F} / \mathcal{R}_{cl}.
$$
\end{definition}

Note that $\mathcal{DQ}$ need not be commutative.
We can realize $\mathcal{DQ}$ as the case $\hbar = 0$ of a family of algebras
parameterized by $\hbar \in \mathbb{C}$ and with $\hbar = 1$ corresponding to $\mathcal{CCR}(P)$.
Based on this we can now define the associated \textit{$\hbar$-deformed relations} to be
\begin{align}
\mathcal{R}_\hbar :=&\langle   \hbar^{n/2} R_0 + \hbar^{(n-1)/2} R_1 + \cdots + 
\hbar^{1/2} R_{n-1} + R_n \, | \, R \in \mathcal{R} \rangle
 \label{define-Rhbar-1x}
\\
= &\langle R_0 + \hbar^{-1/2} R_1 + \cdots + 
\hbar^{-(n-1)/2} R_{n-1} + \hbar^{-n/2} R_n \, | \, R \in \mathcal{R}  \rangle,
   \label{define-Rhbar-2x}
\end{align}
using the notation $R = R_0 + R_1 + \cdots + R_n$ as given above.
Next we define 
$$
    \mathcal{CCR}_\hbar(P) :=  \mathcal{F} / \mathcal{R}_{\hbar}.
$$
The second expression (\ref{define-Rhbar-2x}) has the virtue that the powers of $\hbar^{-1/2}$ are
the degrees of homogeneity of the terms.
On the other hand, in the first expression (\ref{define-Rhbar-1x}) 
each of the homogeneous terms has a coefficient giving its intuitively correct 
degree of `quantumness'.
The expression (\ref{define-Rhbar-1x}) also indicates formally what happens when 
one takes the limit $\hbar \to 0$. 
For $\hbar \ne 0$ the two expressions  (\ref{define-Rhbar-1x}) and  (\ref{define-Rhbar-2x})
are clearly equivalent, but for $\hbar =0$ only the definition (\ref{define-Rhbar-1x}) makes sense.
In physics one considers $\hbar > 0$ to be Planck's constant, 
but here we can take $\hbar \in \mathbb{C}$ to be arbitrary.

We have included $\hbar$ in part to emphasize that this theory has 
semi-classical behavior (more precisely, what happens to $  \mathcal{CCR}_\hbar(P)$
when $\hbar$ tends to zero) as well
as a classical counterpart $\mathcal{DQ}$ (that is, what happens when we put $\hbar$ equal to zero).
However, the developments of  the semi-classical theory and the classical counterpart theory
remain for future research.

Also, it is important to remark that this theory includes both Planck's constant as well as a
Hilbert space where creation and annihilation operators are defined.
These are some of the important characteristics of a quantization relevant to physics.

The Toeplitz algebra, defined as the subalgebra of $\mathrm{End}(\mathcal{P})$
generated by the Toeplitz operators, is also a quantum algebra of interest in itself.

\section{Concluding Remarks}

The point of this note is to develop much of the theory in \cite{tqac} by starting
from a different set of assumptions. 
The inference is that this theory is quite general and probably even more general
than has been worked out so far. 
While non-trivial examples exist in \cite{part2} and \cite{quantum-plane}, there remains
more work to find other applications of this theory. 
Again, the absence of a measure in this approach distinguishes it sharply from other
approaches, such as the coherent state quantization, and so one expects to find
examples of this sort of Toeplitz quantization in settings where other approaches 
do not give results.
I hope that this is not only useful in such mathematical physics contexts, but that
applications of  these ideas from mathematical physics will be useful in the study
of the non-commutative `spaces' of non-commutative geometry (such as quantum groups,
among others) as well as of `spaces' that are even more general.
Also, several open problems were raised during the course of this short note. 
So this is very much a report of work in process.

\subsection*{Acknowledgment}
I thank the organizers of the WGMP XXXII held
in the summer of 2013 for the opportunity to participate in that event.

\end{document}